\documentclass{jfm}
\usepackage{graphicx}
\usepackage{natbib}
\usepackage{float}
\usepackage{subfig}
\usepackage{setspace}
\usepackage{color}
\usepackage{multirow}
\usepackage{amsmath}
\DeclareMathOperator{\tr}{tr}

\newlength{\imw}
\title{Closed-loop separation control
using machine learning}
\author[N. Gautier et al]{N. Gautier$^\dagger$, J.-L. Aider$^\dagger$, T. Duriez$^{\star}$, B.\ R.\ Noack$^\star$, M. Segond$^\ddag$, and M. Abel$^\ddag$}
\affiliation{$^\dagger$ Laboratoire de Physique et M\'ecanique des Milieux H\'et\'erog\`enes (PMMH), UMR7636 CNRS, \\ \'Ecole Sup\'erieure de Physique et Chimie Industrielles de la ville de Paris\\ 10 rue Vauquelin,  75005 Paris, France
\\ 
$^\star$ Institut PPRIME
CNRS - Universit\'e de Poitiers - ENSMA
UPR 3346 \\
D\'epartement Fluides, Thermique, Combustion
CEAT, \\
43, rue de l'A\'erodrome
F-86036 Poitiers cedex, France
\\ 
$^\ddag$ Ambrosys GmbH, \\ 
Albert-Einstein-Str. 1-5, D-14469 Potsdam, Germany }
\date{?; revised ?; accepted ?. - To be entered by editorial office}

\begin{document}
\maketitle
\begin{abstract}
A novel, model free, approach to experimental closed-loop flow control is implemented on a separated flow. Feedback control laws are generated using genetic programming where they are optimized using replication, mutation and cross-over of best performing laws to produce a new generation of candidate control laws. This optimization process is applied automatically to a backward-facing step flow at $Re_h=1350$, controlled by a slotted jet, yielding an effective control law. Convergence criterion are suggested. The law is able to produce effective action even with major changes in the flow state, demonstrating its robustness. The underlying physical mechanisms leveraged by the law are analyzed and discussed. Contrary to traditional periodic forcing of the shear layer, this new control law plays on the physics of the recirculation area downstream the step. While both control actions are fundamentally different they still achieve the same level of effectiveness. Furthermore the new law is also potentially easier and cheaper to implement actuator wise.

\end{abstract}


\section{Introduction}
Flow control is a rapidly evolving interdisciplinary field
comprising many disciplines, like fluid mechanics, technological innovations for sensors and actuators, control theory, optimization and machine learning.
Its  potential engineering applications have an epic proportion,
including aerodynamic of cars, trucks, trains, wind-turbines or gas-turbines, as well as medical equipments or chemical plants, just to name a few.
Flow control is employed to reduce aerodynamic drag  for cars \citep{Beaudoin2008,Kourta2010,Joseph2013}, 
to find alternative lift-off and take-off configurations for aircraft \citep{King2007} and or to improve mixing efficiency \citep{Closkey2002}. 

There have been many successful implementations of passive and active open-loop flow control \citep{Gillieron2010,Joseph2012,Gautier2013upstream}.
Closed-loop control offers great potential for increased robustness and efficiency and is currently the subject of an increasing ongoing research efforts \citep{King2007,Tadmor2004,beaudoin2006drag,Pastoor2008,Marquet2011,Semeraro2011,Gautier2013control}.
In experiments, most closed-loop controls are adaptive \citep{King2007,beaudoin2006drag,Gautier2013control}
and based on slowly varying periodic forcing.
In-time control remains very challenging  
because of the non-linear nature of fluid phenomena.
Most model-based control designs 
are based on a (locally) linear reduced-order model
and ignore frequency cross-talk. 
The low-dimensionality of the model is key for robustness
and online capability in experiments \citep{Noack2011,Cordier2008,Sipp2012}. Only a few control-oriented reduced-order models address frequency cross-talk 
\citep{Luchtenburg2009, Luchtenburg2010}.

The challenges of model-based control design
have lead us to search for model-free control laws
using evolutionary algorithms inspired by nature and evolution
\citep{Wahde2008}. Evolutionary algorithms can indeed be used to search for optimal non-linear control laws. 
They have been successfully used in many disciplines such as bio-informatics, medicine and computer science \citep{Harik1997,Ferreira2001,Hosseini2009}. 
In particular, Genetic Programming (GP) is commonly used to find computer programs that perform a desired task.  
A particular objective of GP is to find a function optimizing a cost functional.
GP is more general than genetic algorithms 
in which parameters of a problem or a given function are optimized.
GP also includes  neural networks.
While these algorithms are commonly 
used in many logistic and pattern recognition tasks,
GP-based control laws represent a paradigm shift 
in experimental closed-loop flow control.
One of the obstacle to application of GP to experimental flow control is that a large number of experiments is required to fulfill the  criterion for statistical convergence. 
Recently,  Duriez et al.\citet{Duriez2013}
developed Machine Learning Control (MLC) 
using genetic programming to find control laws. 
This approach was demonstrated to be surprisingly effective 
when applied on complex dynamical systems 
and even on closed-loop turbulence control in an experiment\citep{Parezanovic2014ftc}. 

The objective of the present study is to use for the first time Machine Learning Control to control a separated flow. It is applied to the recirculation bubble area downstream a backward-facing step (BFS).
In Sect.~\ref{ToC:Experiment}, 
the experiment is described.
MLC is presented in Sect.~\ref{ToC:MLC}.
The MLC-based closed-loop control results are discussed  
and benchmarked against periodic forcing in Sect.~\ref{ToC:Results}.
Sect.~\ref{ToC:Conclusions} summarizes the main finding and provides future directions.

\section{Experimental Setup}
\label{ToC:Experiment}

\subsection{Water tunnel}
Experiments were carried out in a hydrodynamic channel in which the flow is driven by gravity.
The flow is stabilized by divergent and convergent sections separated by honeycombs.  The quality of the main stream can be quantified in terms of flow uniformity and turbulence intensity.  The standard deviation $\sigma$ is computed for the highest free stream velocity featured in our experimental set-up. We obtain $\sigma = 5.9\times 10^{-4}\,\mbox{m.s}^{-1}$ which corresponds to turbulence levels of $\sigma/U_{\infty}=2.3\times 10^{-3}$. For the present experiment the flow velocity is $U_{\infty} =7.3\times 10^{-2}\,\mbox{m.s}^{-1}$  giving a Reynolds number based on step height $Re_h=U_{\infty}h/\nu = 1350$. This Reynolds number was chosen because of the limitations of the injection system. 

\begin{figure}
\centering
\includegraphics[width=0.6\textwidth]{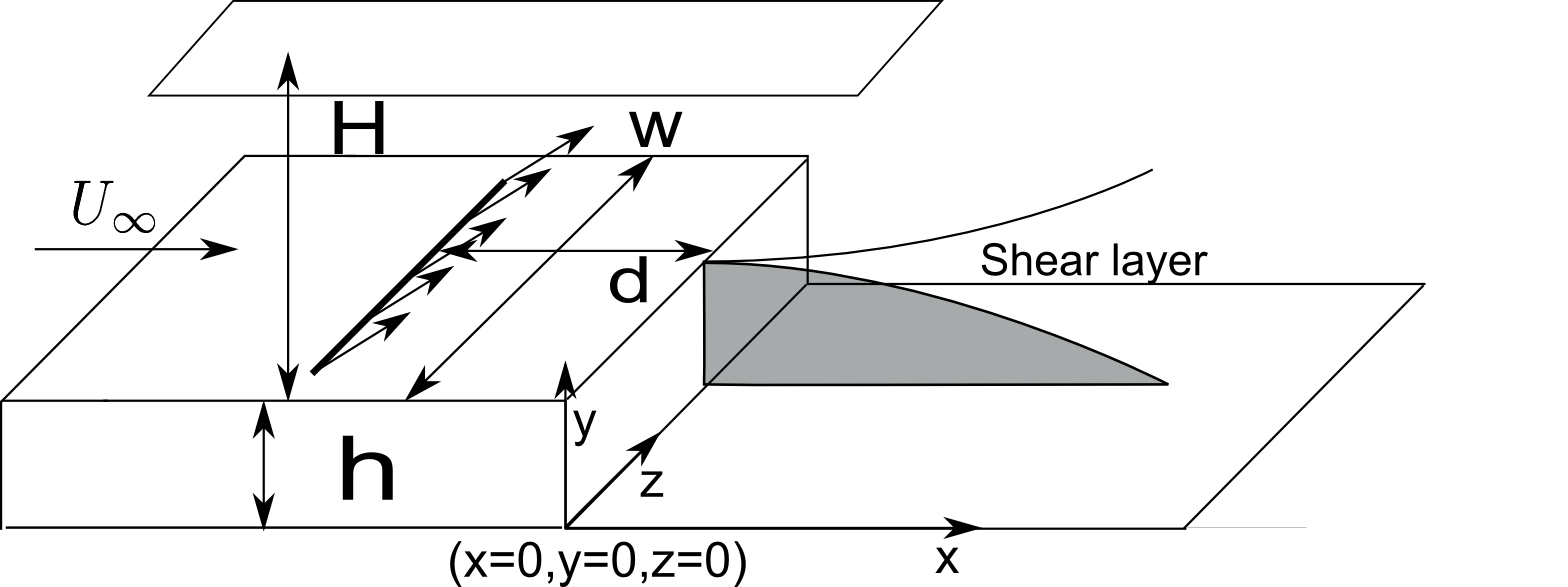}
\caption{Sketch of the BFS geometry, position of the slotted jet and definition of the main parameters.}
\label{fig:dimensions}
\end{figure}

\subsection{Backward-Facing Step geometry}
The BFS is considered a benchmark geometry for the study of separated flows: separation is imposed by a sharp edge creating a strong shear layer susceptible to Kelvin-Helmholtz instability. Upstream perturbations are amplified in the shear layer leading to significant downstream disturbances. This flow has been extensively studied  both numerically and experimentally \citep{Armaly1983,Le1997,JLA2004}. The BFS geometry and the main geometric parameters are shown in figure~\ref{fig:dimensions}. BFS height is $h=1.5\times 10^{-2}\,\mbox{m}$. Channel height is $H=7\times 10^{-2}\,\mbox{m}$ for a channel width $w=15\times 10^{-2}\,\mbox{m}$. The vertical expansion ratio  is $A_y = \frac{H}{h+H} = 0.82$ and the spanwise aspect ratio is $A_z=\frac{w}{h+H}=1.76$. The injection slot is located $d / h =2$ upstream of the step edge. The boundary layer thickness at the step edge is $\delta=1.3h$.

\subsection{Sensor: 2D real-time velocity fields computations}
\begin{figure}
\centering
\subfloat[]{\includegraphics[width=0.75\textwidth,height=0.2\textwidth]{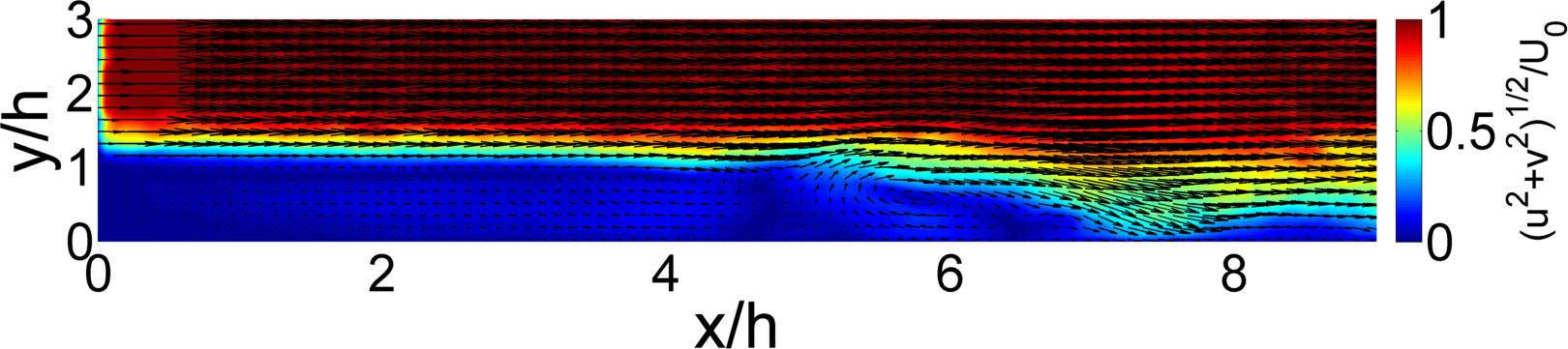}}\\
\subfloat[]{ \includegraphics[width=0.75\textwidth,height=0.2\textwidth]{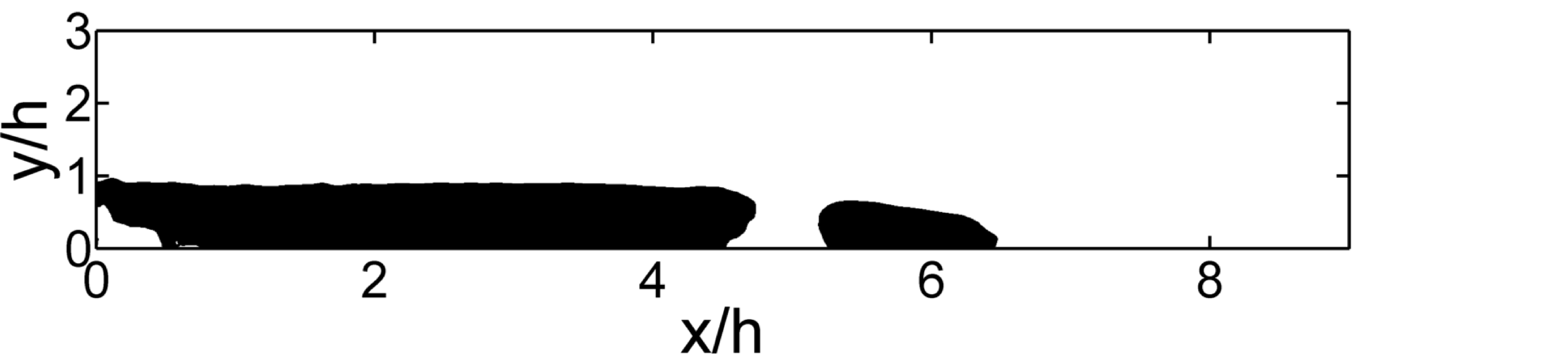}}
\caption{(a) Instantaneous velocity amplitude snapshot. (b) Corresponding instantaneous recirculation area shown in black.}
\label{fig:bubble_snapshot}
\end{figure}

The sensor is built on a real-time computation of the vector fields. The velocity fields are computed based on snapshots from the seeded flow. The seeding particles are 20~$\mu$m neutrally buoyant polyamid particles.  They are illuminated by a laser sheet created by a 2W continuous laser beam operating at $\lambda$~=~532~nm.  Images of the vertical symmetry plane are recorded using a Basler acA 2000-340km 8bit CMOS camera. Velocity fields are computed in real-time on a Gforce GTX 580 graphics card.
The algorithm used to compute the velocity fields is based on a Lukas-Kanade optical flow algorithm called FOLKI developed by \citep{Champagnat2005}. Its offline and online accuracy has been demonstrated and detailed by \citep{Plyer2011,Gautier2013OF}.  Furthermore this acquisition method was successfully used in \citep{Gautier2013control,Leclaire2012,Gautier2014feed}.
The size of the velocity fields is ($17.2\times4.6)\times 10^{-4}\,\mbox{m}^2$. The  time between two snapshots yielding one velocity field is $\delta t = 10\times 10^{-3}\,\mbox{s}$. 42 image pairs are processed per second. Figure \ref{fig:bubble_snapshot} a) shows a typical example of the instantaneous velocity magnitude field downstream the step for the uncontrolled flow.
\\
Recirculation plays a major role in the BFS flow and is overwhelmingly used for flow assessment as well as an objective for the control of flow separation \citep{King2007,Chun1996}.  It has also been shown that the recirculation bubble can be linked to drag (although the relationship is far from trivial) \citep{Dahan2012}. We choose to evaluate the state of the flow through the instantaneous recirculation area, computed from instantaneous velocity fields, as our input. It is a 2D extension of the more common recirculation length evaluated using wall measurements. The recirculation area and recirculation length have been shown to behave the same way by Gautier \& Aider \citep{Gautier2013control}. The normalized instantaneous recirculation area $s(t)$ is computed using equation \eqref{eq:A_r},

\begin{equation}
s(t)=\frac{\int H(-v(t))(x,y)\hspace{1mm}\mathrm{d}x\mathrm{d}y}{A_0}
\label{eq:A_r}
\end{equation}
where H is the Heaviside function and $A_0=1/T\int_0^T A_{uncont}(t)\mathrm{d}t$ is the time-averaged recirculation area for the uncontrolled flow. The figure \ref{fig:bubble_snapshot} b) shows the instantaneous recirculation area corresponding to the instantaneous velocity field shown on figure \ref{fig:bubble_snapshot} b) and computed using equation \eqref{eq:A_r}.\\

\subsection{Actuator}
Actuation is provided using upstream injection through a spanwise slotted jet as shown in figure \ref{fig:dimensions}. The angle between the jet axis and the wall is 45$^o$. The jet flow is induced using a pressurized water tank. It enters a plenum and goes through a volume of glass beads designed to homogenize the incoming flow. The jet amplitude $U_j$ is controlled by changing tank pressure. Because channel pressure is higher than atmospheric pressure this allows us to provide both blowing and suction. Maximum actuation frequency $f_a$ is about 2Hz. To achieve closed-loop control, the control value $b={U_j}/U_{max}$ ($U_{max}$ being the maximum jet velocity) is computed as a function of the sensor value $s$ inside a Labview project. The specific control laws are derived using machine learning control.

\section{Machine Learning Control}
\label{ToC:MLC}
\begin{figure}
\centering
\includegraphics[width=0.55\textwidth]{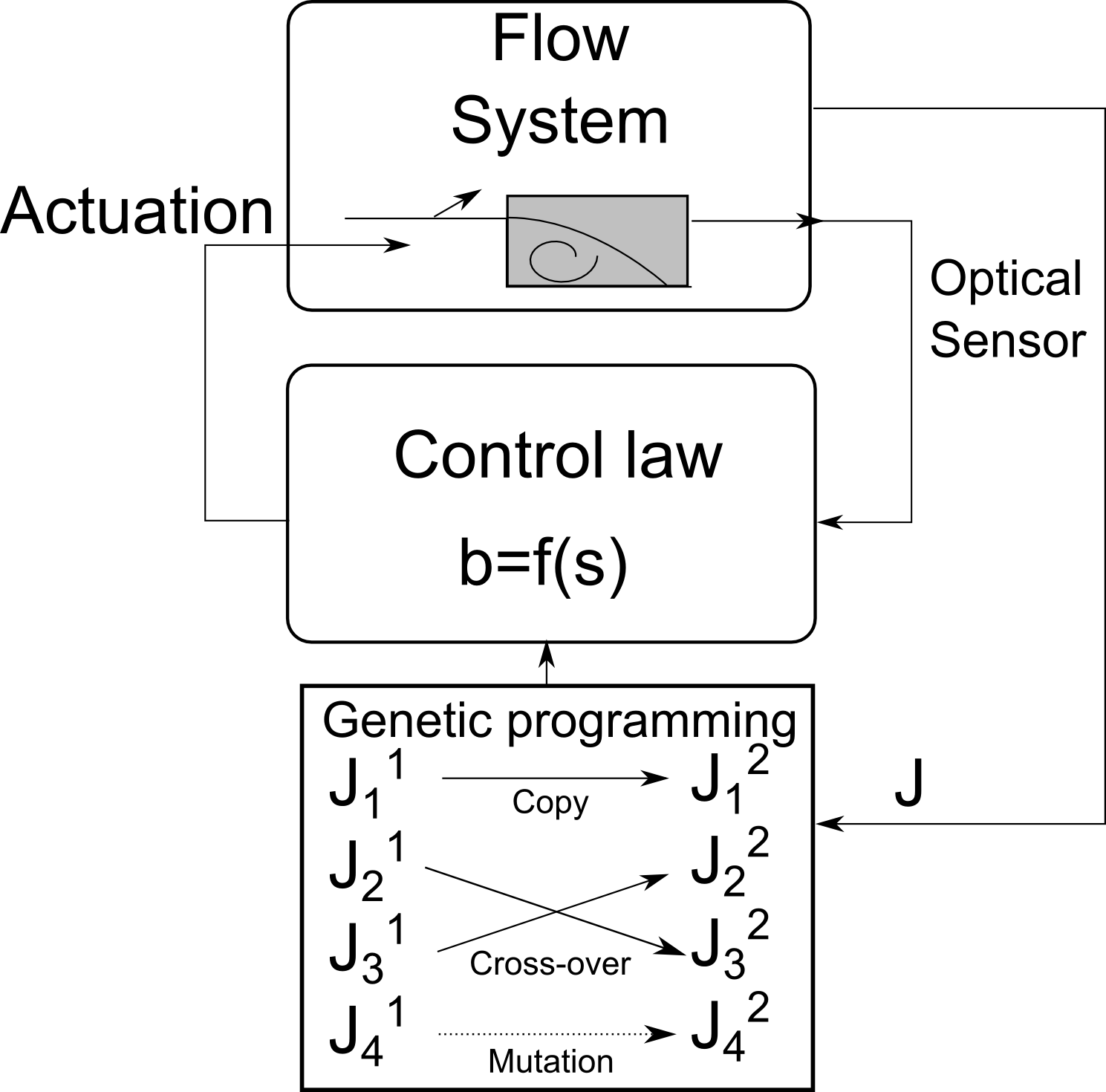}
\caption{Control loop featuring genetic programming. Control laws $b(s)$ are evaluated by the flow system. This is done over several generations of individuals. New generations are generated by replication, cross-over and mutation $J_i^n$ refers to the $i^{th}$ individual of generation $n$.}
\label{fig:genetic_programming_sketch}
\end{figure}

Machine learning control is a generic, model free, approach to control of non-linear systems. Control laws are optimized with regard to a problem specific objective function using genetic programming \citep{koza}. A first generation of control laws candidates $b_i^1(s)$ , called individuals ($b_i^1(s)$ is the $i^{th}$ individual of the $1^{st}$ generation), is randomly generated by combining user defined functions, constants and the sensor value $s$. Each individual is evaluated yielding a value for the cost function $J$. A new population $b_i^2$ is then generated by evolving the first generation. The procedure is iterated until either a known global minimum of $J$ is reached or the evolution is stalled. This process is resumed in figure~\ref{fig:genetic_programming_sketch}.

\subsection{Population generation}
The first individuals $b_i^1$ are generated as expression trees made of user-defined nodes (see Appendix~\ref{ToC:AppA}). These nodes are functions ($\sin$, $\cos$, $\exp$, $\log$, $\tanh$), basic operations ($+,-,\times,/$), constants and the sensor input $s(t)$. The root of the expression tree, i.e. the value returned by the function it defines, is the control value. To build the expression tree, a recursive algorithm is used: a first node is chosen, then for each argument this node can accept, new nodes are added randomly until all terminal nodes do not accept any arguments (constants or sensor). The algorithm is made so that the first generation contains expression trees of different depth and density to ensure diversity in the population. Furthermore all individuals are different.

The number of individuals inside a generation has a strong influence on the MLC process. While a large number of individuals will certainly lengthen the total time of the experiment, it will also ensure a higher diversity which is known to be a key parameter of all evolutive algorithms. In the present study, each generation is made of 500 individuals. This number of individuals is a good compromise between performance and testing time. It has proven to be enough to converge on most single input/single output problem, and is still manageable in terms of total experimental time.

\subsection{Evaluation}
Expression trees can be easily written as LISP expressions. The evaluation is done by translating the LISP expression (for example  $(+ (\sin s) (/ 2.343 s))$) into a control law inside the software responsible for the real-time closed-loop control ($b=\sin(s+(2.343 / s)$). The numerical value used to grade each individual is the cost function $J$ linked to the control problem. In our case the goal is to reduce the recirculation area over the evaluation time with a penalization by the  energy used for the actuation:

\begin{equation}
J=\left( \langle s \rangle + w \langle |b| \rangle^2 \right)>0
\label{eq:cost_function}
\end{equation}

where $T$ is the evaluation time. The first component, quantifies the state of the flow and integrates the sensor $s(t)$ during the evaluation time. $s(t)$ is normalized by the time-averaged uncontrolled recirculation area $A_0$. Normalization is important as it allows corrections taking into account variations in the flow conditions (i.e. temperature variations, flow rate drifts). $A_0$ is recomputed every 250 individuals to compensate for any drifts. The second component, weighted by $w$, is tied to actuation energy and is normalized by the maximum jet velocity $U_{max}$. In the following, we choose $w=3/2$, to strongly penalize high actuation costs. 

\begin{figure}
\centering
\includegraphics[width=0.55\textwidth]{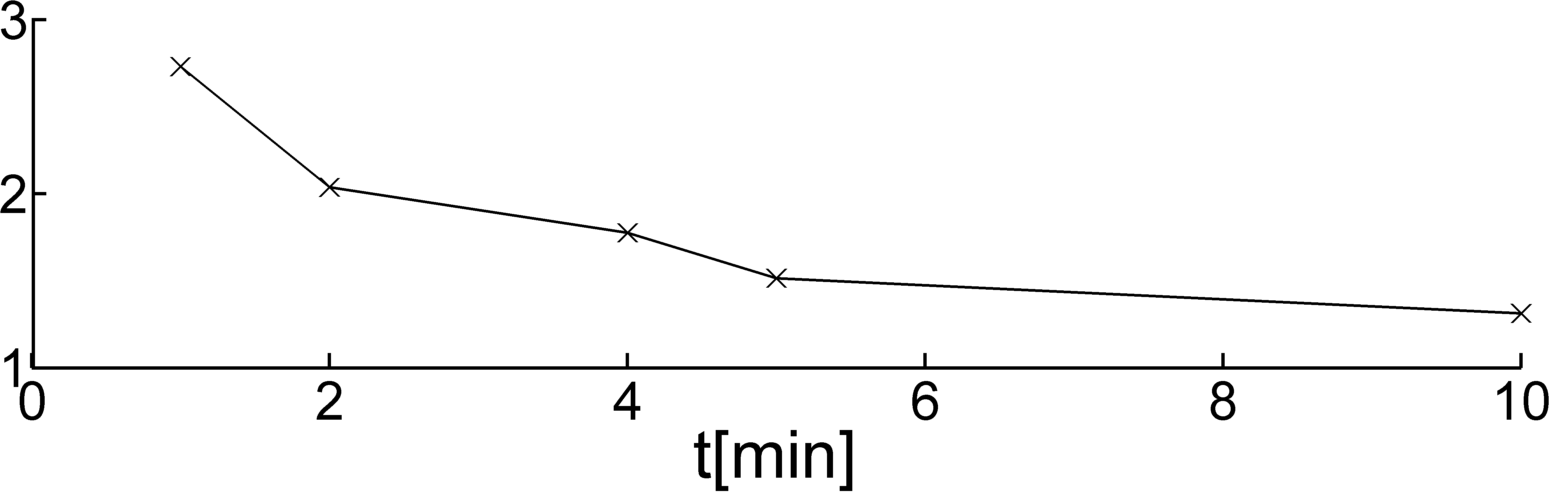}
\caption{Variations of of the mean recirculation area (\%) over ten evaluations for different evaluation periods.}
\label{fig:eval_std}
\end{figure}

The evaluation time is also a key parameter as it will determine how long the whole experiment will last. Figure~\ref{fig:eval_std} shows that a one minute evaluation is enough to get significant statistics for an evaluation of $J$ good enough to discriminate individuals by performance. As 
time is spent refilling the jet supply tank, the time between two evaluations varies and can reach two minutes. Approximately 1000 individuals, i.e.  two generations, are evaluated over 24 hours.   \\

Because control laws can be constants or give a constant response, each control law is pre-evaluated before it is applied to the flow. If it is found to saturate the actuator, it is assigned a very high cost. This step takes a few milliseconds and is done to ensure faster convergence by discarding uninteresting functions. Because of the random nature of the first generation most individuals saturate the actuator.

\subsection{Breeding of subsequent generations and stop criteria}
Once every individual of the current generation has been evaluated, they are sorted by their cost function value $J$. The five best individuals are evaluated again, the cost values are averaged and the population is sorted again. This re-evaluation procedure is repeated five times to ensure that the value of the best individuals is reliable. The individuals of the next generation are then produced through 3 
 different processes. 
Mechanisms are based on a tournament process: 7 individuals are randomly chosen, the individual elected to enter a breeding process is the one with the lowest cost function value. This ensures that the best individuals inside a generation will be used a lot, while less performing individuals still have a chance to be part of the next generation. Individuals selected this way will then be either replicated, mutated or crossed to generate the individuals of the next generation. The probabilities of replication, mutation and crossover are respectively $10\%$, $20\%$ and $70\%$. This new generation is then evaluated and the whole process is iterated. 
The MLC process can stop for two reasons. The first one is when $J$ reaches $0$ which in general does not occur. One can stop the process when the best values of $J$ over the population stop improving over several generations.

\section{Results}
\label{ToC:Results}
Machine learning control described in \S \ref{ToC:MLC} 
has been applied to the backward-facing step plant presented in \S \ref{ToC:Experiment}.
Convergence of MLC generations is analyzed in \S \ref{ToC:Convergence}.
The best control law of the final generation 
is presented in \S \ref{ToC:BestControlLaw}.
This MLC control law is compared 
with open-loop forcing (\S \ref{ToC:PeriodicForcing})
and tested for robustness with respect to the Reynolds number in \S \ref{ToC:Robustness}.

\subsection{Convergence of machine learning control}
\label{ToC:Convergence}
\begin{figure}
\centering
\includegraphics[width=0.70\textwidth]{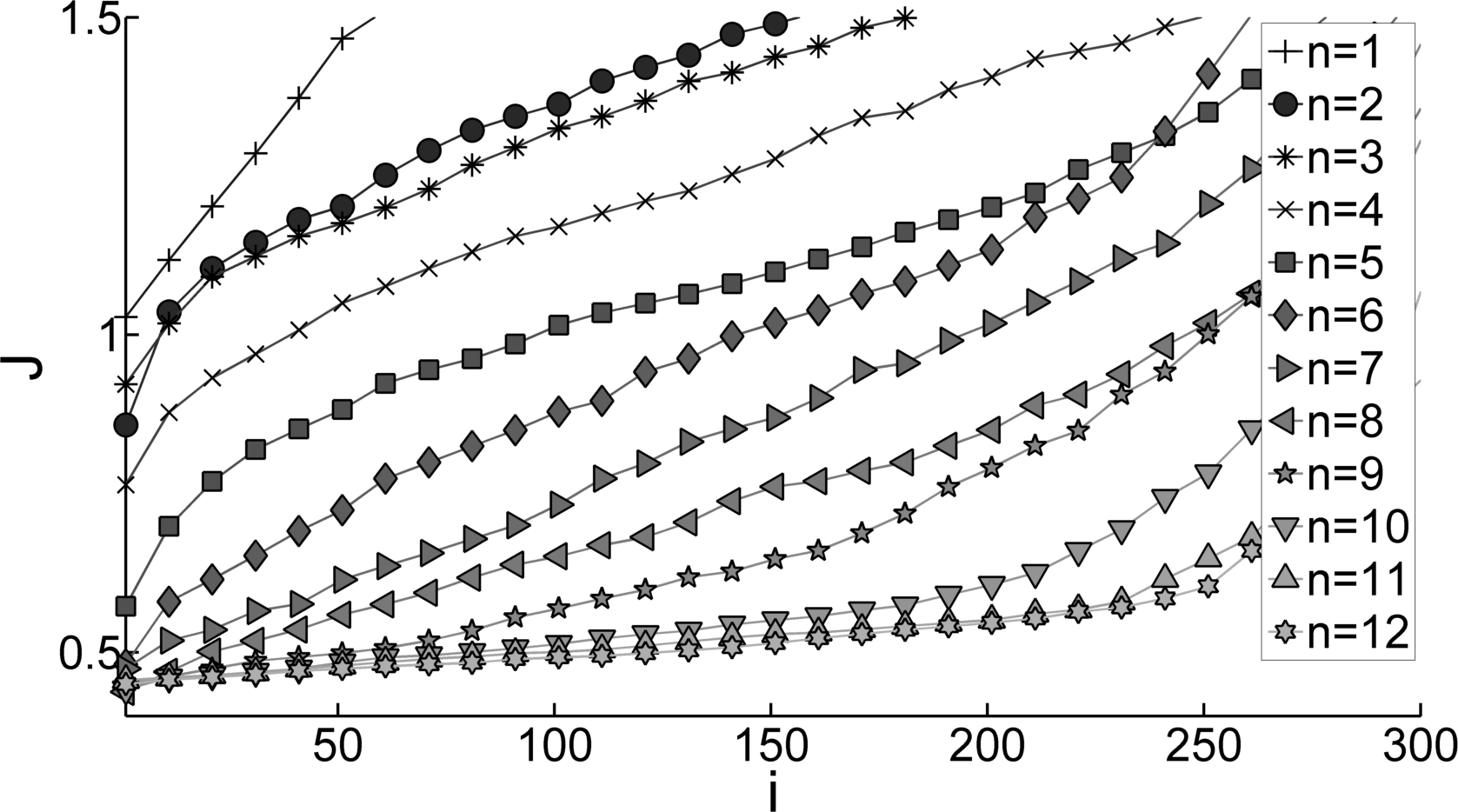}
\caption{Cost functions of the first 300 individuals in all twelve generations.}
\label{Fig:fitness_vs_generations}
\end{figure}

The evolution of the cost function with increasing number of individuals 
is shown in Figure \ref{Fig:fitness_vs_generations} for all twelve generations.
All random control laws of the first generation 
are seen to be ineffective, 
i.e.\ produce only cost functions 
which are worse than the uncontrolled flows ($J_i^1 >1$ for all $i$).
A few effective control laws can be seen as soon as the second generation.
The slope of the cost function $J_i$ as a function of the  
 index $i$ is improved by MLC.
All subsequent generations perform better than the previous one, i.e.\ $J_i^{n+1} < J_i^n$. 
After the $9^{th}$ generation, the performance of the best individuals appear to converge.

\begin{figure}
\centering
\includegraphics[width=0.70\textwidth]{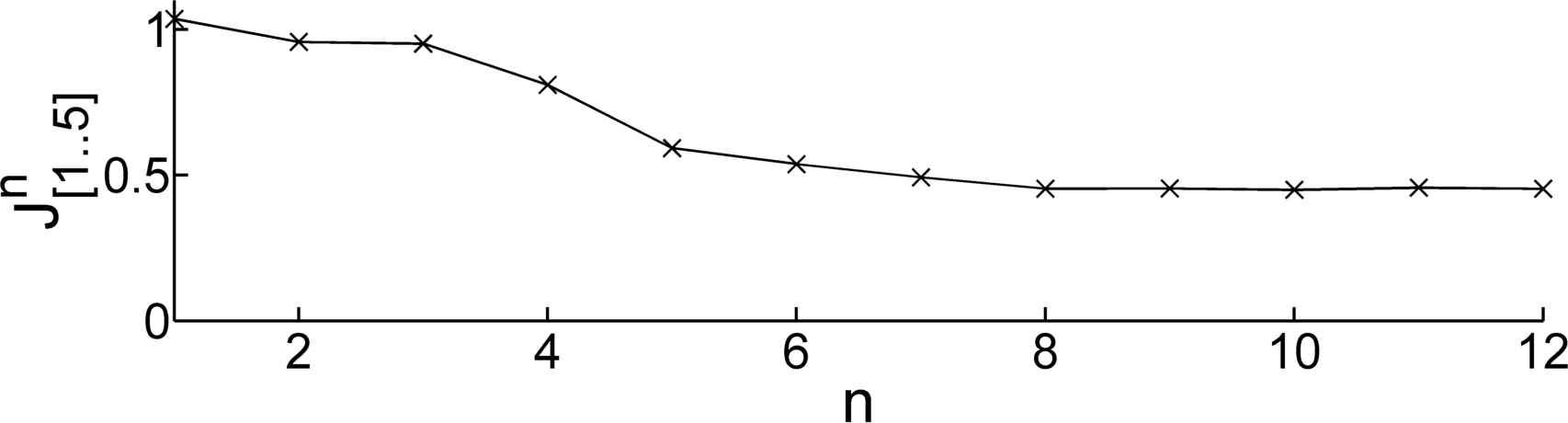}
\caption{Cost functions averaged over the five best individuals for each generation}
\label{Fig:Convergence}
\end{figure}

The average of the best five control laws 
$J_{[1..5]}^n := \left ( 
  J_1^n+ J_2^n + J_3^n + J_4^n + J_5^n
\right) / 5$
is shown in  figure \ref{Fig:Convergence} for each generation.
Convergence is reached after the $8^{th}$ generation.
A good termination criterion appears to be to stop the iteration 
once the average of the cost function for the first 5 individuals 
no longer improves.
As the number of generations increases 
the first 5 control laws become very similar
and averaging over them is a more robust measure than taking just the best one.

\subsection{Analysis of the best control law obtained by MLC}
\label{ToC:BestControlLaw}
\begin{figure}
\centering
\includegraphics[width=0.65\textwidth]{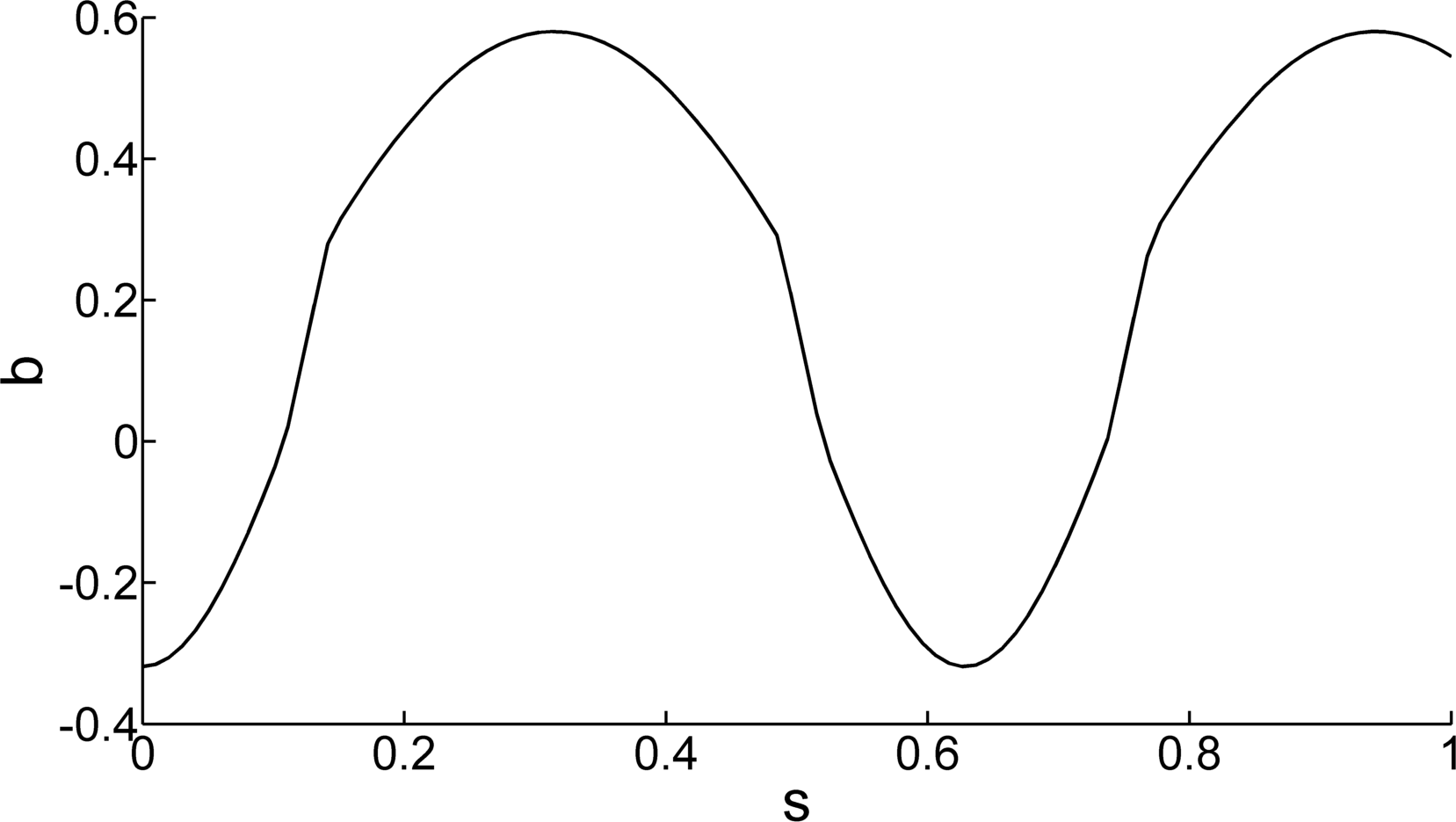}
\caption{Graph of the best control law $b=K(s)$ 
obtained after 12 generations.}
\label{Fig:control_law}
\end{figure}

The control law $b=K(s)$ 
has a complex mathematical expression. 
Yet, the graph of the best control law for the final generation
has a simple structure as shown on figure \ref{Fig:control_law}
for $0\le s \le 1$.
Indeed, the controlled normalized recirculation area 
is by definition positive, $s \ge 0$.
Experimentally, the instantaneous recirculation area
associated with the best MLC law
is found to  be always smaller than the averaged uncontrolled region
resulting in $s \le 1$.  Note that the control law leads to a combination of blowing and suction as a function of $s$.

The actuation command $b$ has an interesting non-monotonous dependency²
on $s$
with two similar maxima $b \approx 0.6$  (injection)
and two similar minima $b \approx 0.3$ (suction).
Near uncontrolled values for the recirculation zone ($s\sim 1$), 
large jet injection reduces the area.
This injection decreases with $s$ until 
suction sets in at intermediate values ($s \sim 0.73$).
In the post-transient regime ($0.12 < s < 0.32$),
injection increases with recirculation area.
For $s < 0.12$, suction sets in.
Most of the time, injection $b \approx 0.5$ occurs. 
During short periods with low recirculation zones, 
suction sets in or, at minimum, injection is significantly reduced.

\begin{figure}
\centering
\subfloat[]{\includegraphics[width=.65\textwidth]{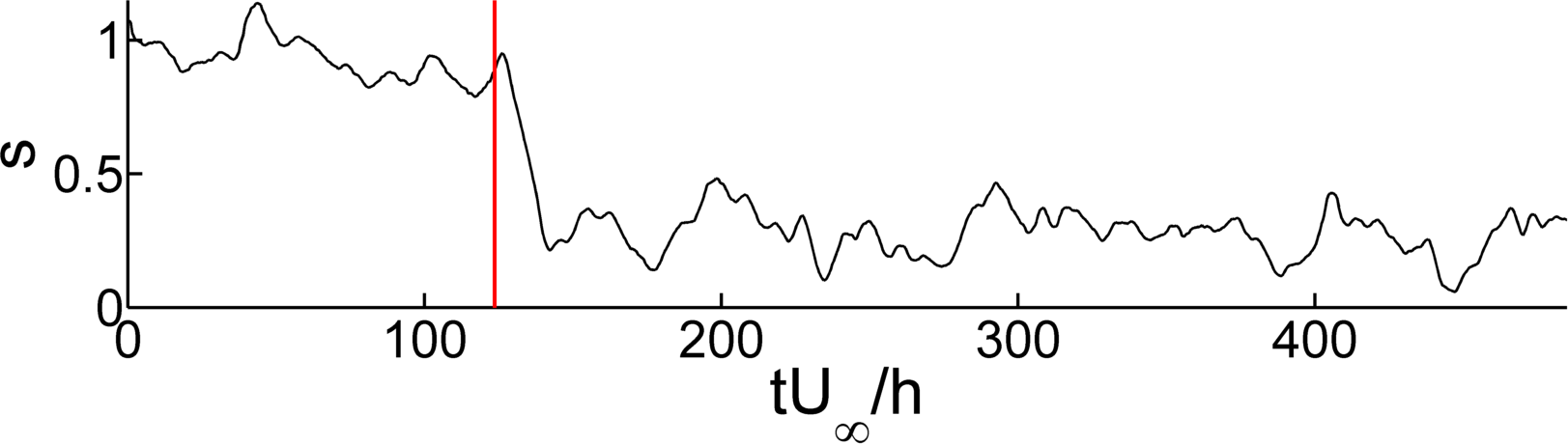}\label{Fig:control_law_output}}\\
\subfloat[]{\includegraphics[width=.65\textwidth]{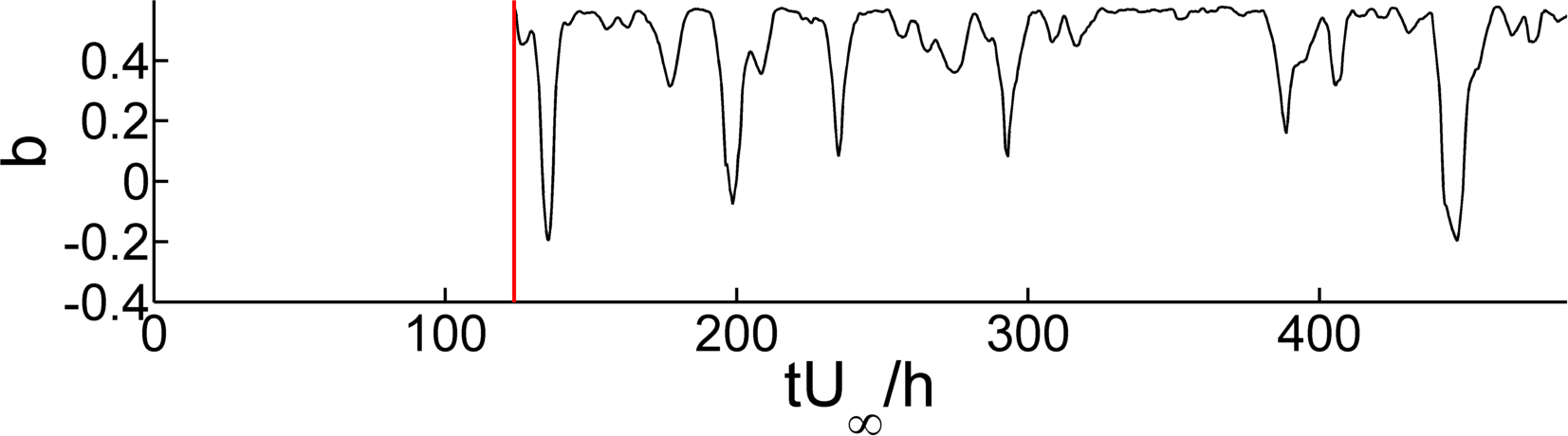}\label{Fig:control_law_action}}
\caption{(a) System response $s$ to the control law, the vertical line shows when control starts. (b) Corresponding actuation $b$.}
\end{figure}

It is interesting to look at the time-series of $s(t)$ (figure \ref{Fig:control_law_output}) and the corresponding actuation amplitude $b(t)$ (figure \ref{Fig:control_law_action}) for the best closed-loop control law. Once the control starts at time $t = 25$ s (vertical red line) 
the recirculation area is quickly and strongly decreased down to 20 \%, 
 corresponding to a $80\% $ reduction on average. For  $0<s<0.3$,  actuation $b$ is roughly a linear function of $s$
(see figure \ref{Fig:control_law}). The figure \ref{Fig:control_law_action} shows that the actuation is indeed a succession of short period of suction followed by a longer period of blowing ($b \approx 0.45$).

\begin{figure}
\centering
\includegraphics[width=0.75\textwidth]{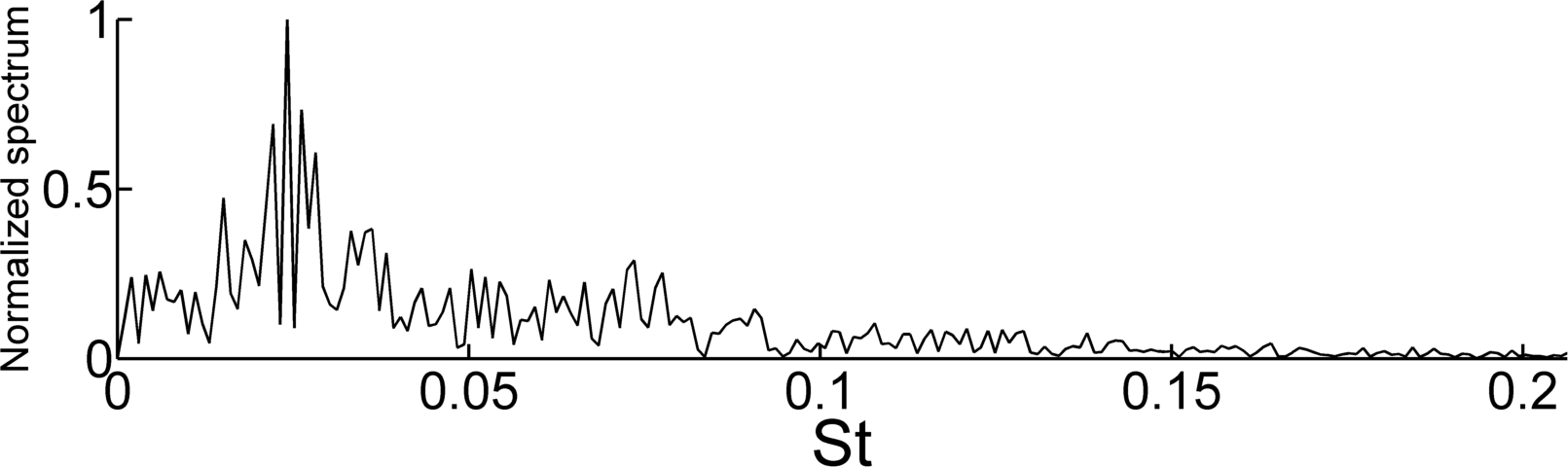}
\caption{Normalized frequency spectrum obtained by Fourier transform of the actuation signal.}
\label{fig:frequency_analysis}
\end{figure}

This feedback loop creates oscillations at 0.1 Hz, 
as observed in figures \ref{Fig:control_law_output} and \ref{Fig:control_law_action} and confirmed by the frequency analysis for the actuation signal  shown in figure \ref{fig:frequency_analysis}. This frequency is indeed close to the flapping frequency of the recirculation bubble which is typically an order of magnitude lower than the shear layer shedding frequency \citep{Cicca2001}, which is close to 1 Hz at this Reynolds number.

The 0.1 Hz feedback dynamics is probably triggered
by the choice of our input, 
the instantaneous recirculation area,
and the natural flapping frequency.
The periodic events of reduced injection
or low suction are awarded by the cost function 
which penalizes the actuation.
\\
It is important to notice that the low frequency nature of this actuation enables "slow" actuators to positively affect high Reynolds number flows. It may remove strong constraints on the actuator in an industrial settings which usually deale with high frequency vortex sheddings (typically a few hundreds Hz for full-scale automotive aerodynamics). In addition, it has been shown that recirculation area and recirculation length behave in the same way \citep{Gautier2013control}. Wall pressure sensors could be used to evaluate recirculation length in real-time \citep{King2007} which could be used as an input to this new control law, making realistic applications viable .


\begin{figure}
\centering
\subfloat[]{\includegraphics[width=0.75\textwidth]{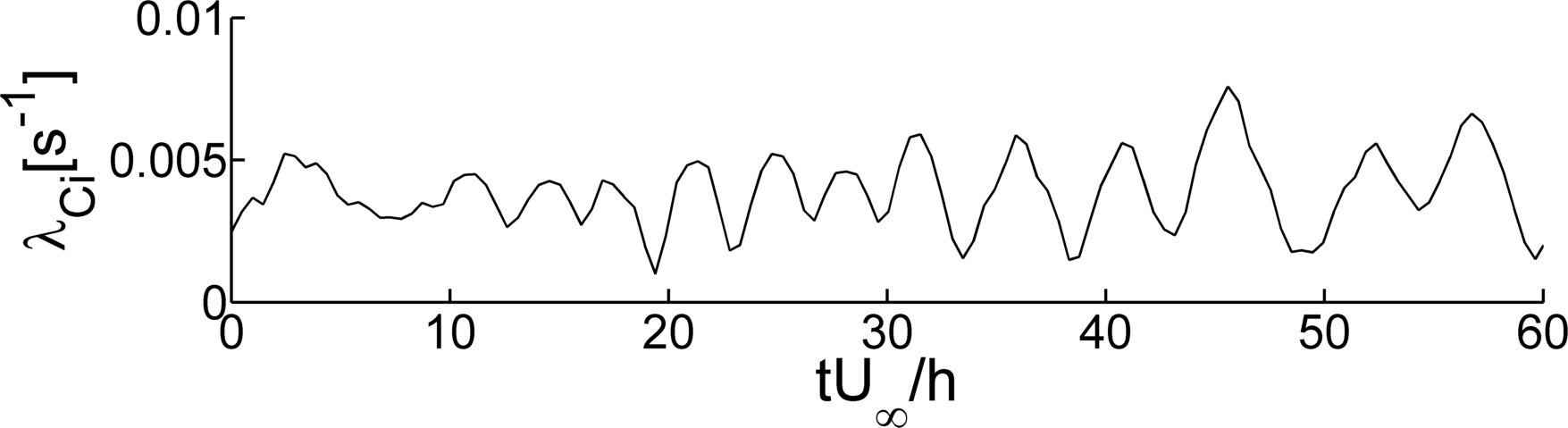}\label{fig:natural_shedding_signal}}\\
\subfloat[]{\includegraphics[width=0.75\textwidth]{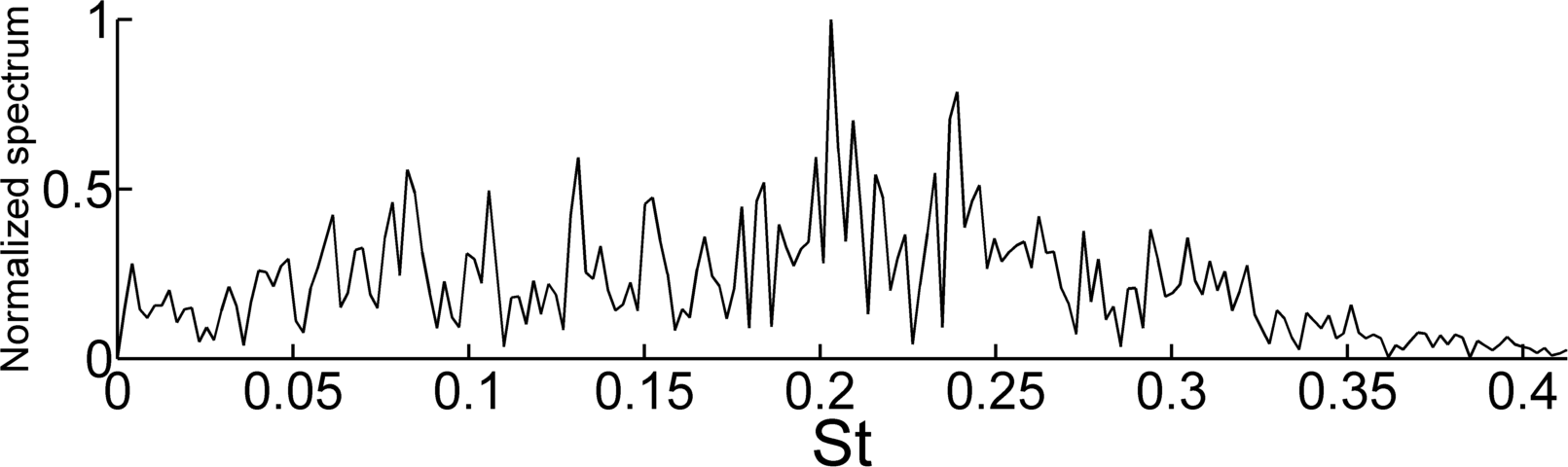}\label{fig:natural_shedding_spectrum}}
\label{fig:natural_shedding}
\caption{(a) $\lambda_{Ci}(t)$ signal over 60 seconds, uncontrolled flow. (b) Normalized frequency spectrum showing a peak at 0.97 Hz.}
\end{figure}

\subsection{Comparison to periodic forcing}
\label{ToC:PeriodicForcing}
Pulsing jet injection at the natural shedding frequency is an effective way of reducing recirculation area \citep{Chun1996,Pastoor2008,Gautier2013upstream} and is then a natural benchmark for MLC.
We choose a periodic forcing at the Kelvin-Helmholtz frequency with a duty cycle of 50\%.
 An effective way of computing the natural shedding frequency is to compute the swirling strength criterion $\lambda_{ci}(s^{-1})$.  This criterion was first introduced and subsequently improved by \citep{Chong1990,Zhou1999}. It was also recently used as an input in closed-loop flow control experiments \citep{Gautier2013control}. For 2D data $\lambda_{Ci}$ can be computed quickly and efficiently following equation~\eqref{eq:lCi},

\begin{equation}
\lambda_{Ci}=\frac{1}{2}\sqrt{4 \det(\nabla \bf{u})- \tr(\nabla \bf{u})^2}
\label{eq:lCi}
\end{equation}

when such a quantity is real, else $\lambda_{Ci}=0$.  
The shedding frequency is obtained by spatially averaging $\lambda_{Ci}$ in the vertical direction at $x=5 h$. The sampling frequency is $f_s =$ 10Hz. This is equivalent to counting vortices as they pass through an imaginary vertical line at $x=5h$. Figure \ref{fig:natural_shedding_signal} shows the corresponding scalar over 60 seconds. Figure~\ref{fig:natural_shedding_spectrum} shows the frequency spectrum obtained by Fourier transform. The natural shedding frequency is well defined and close to 1 Hz. This gives us the frequency for periodic forcing.\\

\begin{figure}
\centering
\subfloat[]{\includegraphics[width=0.75\textwidth]{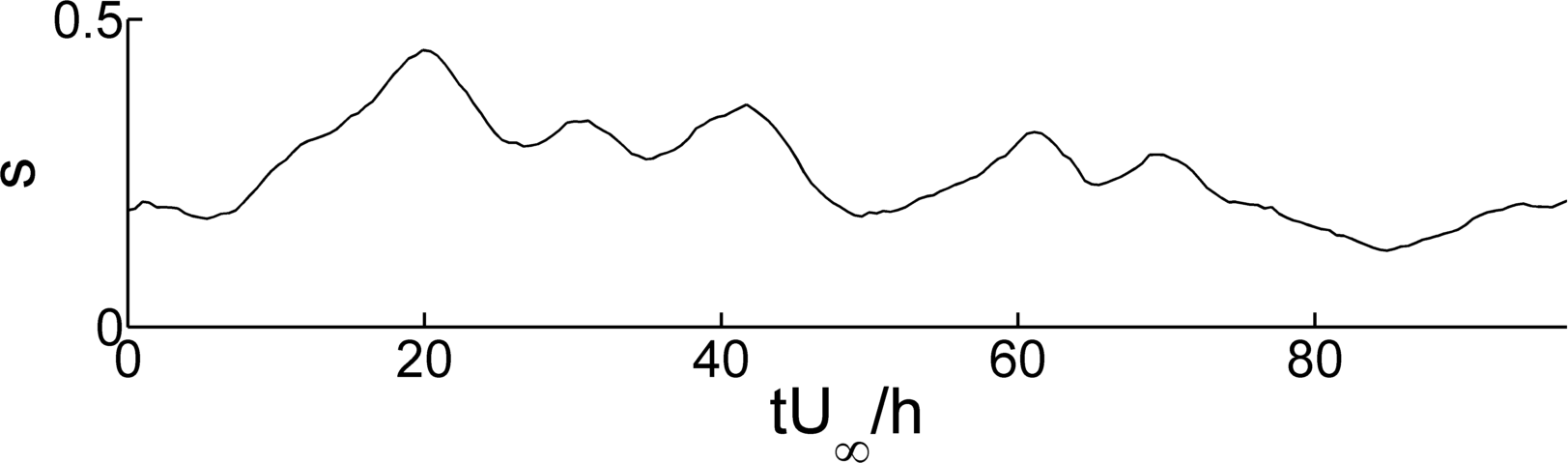}\label{Fig:control_law_freq_output}}\\
\subfloat[]{\includegraphics[width=0.75\textwidth]{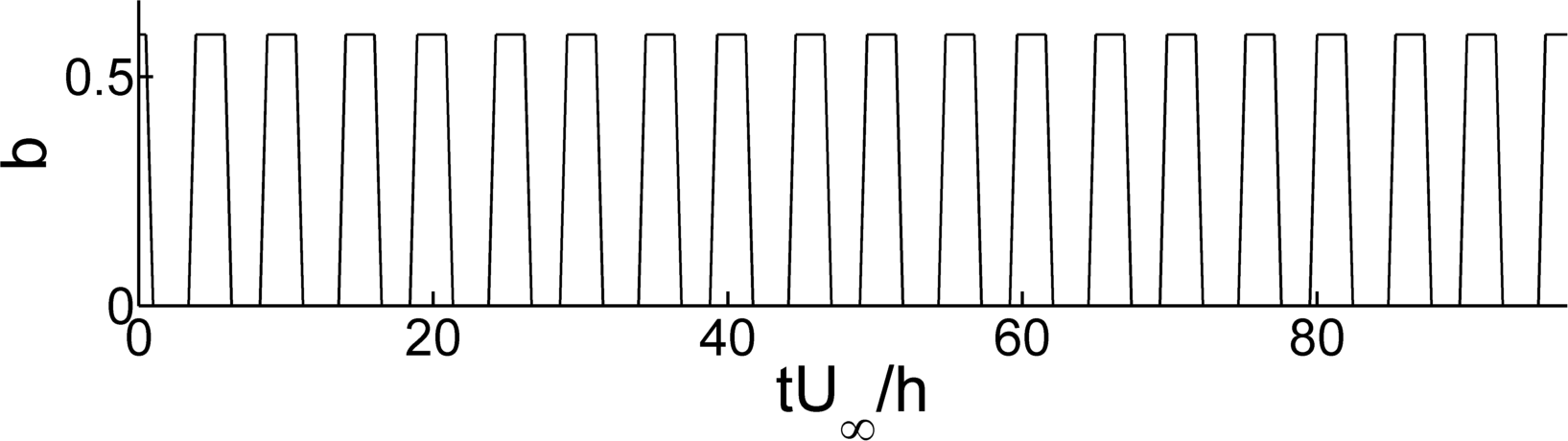}\label{Fig:control_law_freq_action}}
\caption{(a) System response to periodic forcing over 20 seconds. (b) Corresponding actuation.}
\end{figure}

Figure \ref{Fig:control_law_freq_output} shows the reduction of the recirculation area using periodic forcing 
(figure \ref{Fig:control_law_freq_action}). 
Control is effective in reducing the recirculation. 
The cost function for this control law is $J_{periodic}=0.423$ which 	
is quite similar to the one found for MLC ($J_{MLC}=0.419$). 
The MLC based law still performs slightly better.
Intriguingly, similar performances are achieved but
with quite different dynamics and frequencies.
The periodic forcing excites the Kelvin-Helmholtz frequencies at $1$ Hz
while MLC exploits the flapping frequency around $0.1$ Hz.

\subsection{Robustness}
\label{ToC:Robustness}
The control law was tested for various Reynolds numbers in order to test its robustness. Table \ref{tab:robustness} shows the resulting cost functions. The cost function does not increase by more than 20\%  while Reynolds number changes by a factor 2. The cost function for the open loop forcing is also shown. This open loop forcing is done at the optimal frequency for $Re_h=1350$ and is not changed with Reynolds number. Because it is tied to the recirculation area MLC based control law adapts to changes in operating conditions, ensuring consistent, reliable performances. Because natural frequency changes with free-stream velocity, the open-loop control shows poor performance at different Reynolds numbers. Performance variations for the MLC control law are due to changes in jet to cross-flow momentum ratio. To further improve robustness the control law should be amended in the following way: $b=f(Re_h)K(s)$ where $f$ is a function to be determined. 
\begin{table}
\centering 
\begin{tabular}{c c c c}
$Re_h$ &  900 & 1350 & 1800   \\ \hline 
$J_{closed-loop}$  & 0.40 & 0.42 & 0.49 \\ \hline 
$J_{open-loop}$ & 0.91 & 0.42 & 0.98 
\end{tabular}
\caption{Cost function of MLC at different Reynolds numbers.}
\label{tab:robustness}
\end{table}


\section{Conclusion}
\label{ToC:Conclusions}
Machine learning control has been used to determine a cost effective control law minimizing recirculation on a backward-facing step flow. During the twelve generations needed to converge towards the control law the population has evolved toward solutions better fitting the problem as graded by the cost function value. The process is stopped when no amelioration can be foreseen, judging by the statistical values returned by the MLC algorithm. As no convergence can be proven, there is no guarantee this control is optimal, however the nature of the cost function allows to judge the performance of the solution and whether actuation can be rated as effective.
\\
Without deriving a model for the input-output system, MLC is able to converge on an efficient and robust control law linking a real-time measure of recirculation area to actuation value. Though the design of the experiment is kept at its simplest, an 80\% reduction of the recirculation area has been achieved. Genetic programming which is at the core of MLC has proven to be efficient at resolving multi input/multi output problems. Thus adding more freedom to the algorithm, by adding control outputs and sensors inputs, will increase the number of mechanisms MLC has access to in order to reduce the cost function. 
\\
 Robustness can be increased or decreased by closed-loop laws. It is demonstrated that the control law designed by MLC for a given Reynolds number stays efficient in other operative conditions. Robustness would be reinforced by integrating a condition change during evaluation, though it would lengthen overall evaluation time.  
\\
This novel approach of experimental flow control competes with other approaches in terms of efficiency and robustness. Being model-free and capable of producing virtually any kind of control law (linear, non-linear, with history of sensors and actuators) it can be used in a systematic fashion, with a known time consumption on the plant. This enables the method to be used on flows with a specific geometry which has not been thoroughly investigated, such as a detailed vehicle model or a turbine geometry.

\section{Acknowledgements}
NG and JLA wish to thank the DGA for their support, TD and BN acknowledge funding by the French ANR (Chaire d'Excellence TUCOROM and SEPACODE), MS and MA acknowledge the support of the LINC project (no. 289447) funded by ECs Marie-Curie ITN program (FP7-PEOPLE-2011-ITN).

\bibliographystyle{jfm}
\bibliography{Bibliography}
\clearpage
\appendix
\section{}

\label{ToC:AppA}
In this appendix we provide further explanations about how control laws are translated from expression trees (\S \ref{ToC:exptrees}) and how the mutation and crossover operations are performed (\S \ref{ToC:GPoperations}).

\subsection{Control laws and expression trees.}
\label{ToC:exptrees}
\begin{figure}
\centering
\includegraphics[width=0.3\textwidth]{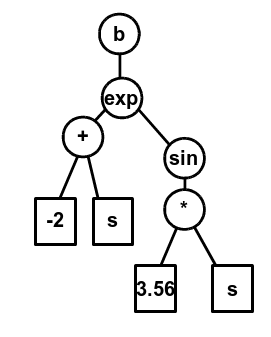}
\caption{A typical expression tree.}
\label{fig:exptree}
\end{figure}

An expression-tree can be viewed in a graphical way as a tree-like representation of the function under consideration as in figure~\ref{fig:exptree} with nodes (round shapes) representing user-defined functions and leaves (square shapes) representing the constants and inputs of the function. The root of the tree (the top node) is the function output.
This tree can also be described by a LISP expression. A LISP expression is easily generated and manipulated from a computational point of view. For instance the function $b(s)=\exp(-2s + sin(3.56*s))$ which is represented by the tree in figure~\ref{fig:exptree} is represented by the LISP expression $ (\exp (+ (* -2 s) (\sin (* 3.56 s)))$. The fact that the operator comes first allows to generate, evaluate and manipulate the individual with recursive functions.

\subsection{Genetic programming operations on expression trees.}
\label{ToC:GPoperations}

\begin{figure}
\centering
\includegraphics[width=0.55\textwidth]{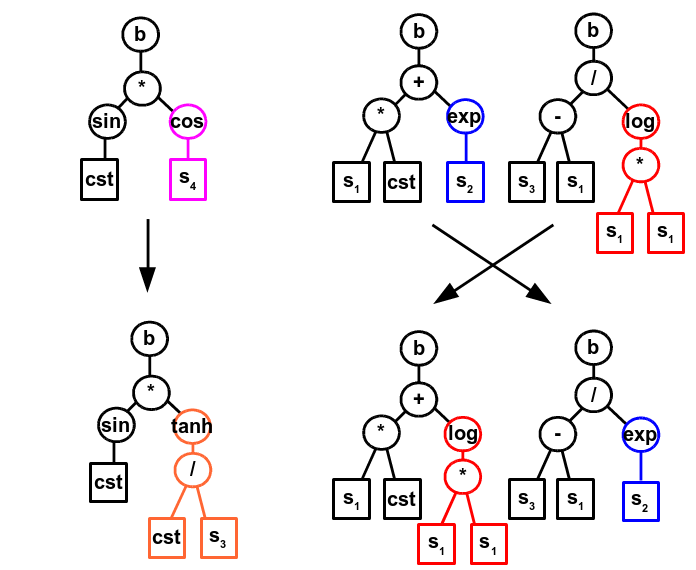}
\caption{Left: a possible mutation of an individual. Right: a possible crossover between two individuals}
\label{fig:GPop}
\end{figure}

The figure~\ref{fig:GPop} illustrates how the operations of mutation and crossover are performed.
 
The mutation operations (left) are performed by selecting a node, erasing the node and its subtree and growing a new subtree randomly. Part of the information contained in the individual is kept while new information is allowed to enter the population. The mutation operation increases the diversity and is responsible for exploring the search space with larger steps. 
The crossover operation (right) consists in selecting one node in each of the two individuals under consideration. The nodes and their subtrees are then exchanged. No new content is brought in  but combinations of operations from good individuals (they both won a 7 contestants tournament) are tested together. The crossover is responsible for exploring the search space around individuals that are performing correctly.
By adjusting the crossover and mutation probabilities, it is possible to adjust the genetic programming way of converging. A high rate of mutation will explore more of the search space while a high rate of crossover will converge faster around detected minima, whether global or local.

\end{document}